\documentstyle[12pt,fleqn]{article}
\def\be{\begin{equation}}
\def\ee{\end{equation}}
\def\ba{\begin{eqnarray}}
\def\ea{\end{eqnarray}}

\def\fun#1#2{\lower3.6pt\vbox{\baselineskip0pt\lineskip.9pt
        \ialign{$\mathsurround=0pt#1\hfill##\hfil$\crcr#2\crcr\sim\crcr}}}
\def\re#1{{$^{\ref{#1}}$}}

\def\m{{\cal M}}

\def\ApJ{{\it Astrophys.~J.} }
\def\numu{{\nu}_{\mu}}
\def\nuee{{\nu}_e}
\begin{document}
\begin{titlepage}
\null\vspace{-62pt}
\begin{flushright} DOE--ER--40682--26\\
\\
Revised Version\\
\\
To appear in Phys.Rev.D\\.
\end{flushright}
\vspace{0.5in}
\centerline{\large \bf{Quasar Production:Topological Defect Formation due to
a}}
\centerline{\large \bf{Phase Transition linked with Massive Neutrinos}}
\vspace{.5in}
\centerline{Anupam Singh}
\vspace{.4in}
\centerline{{\it Physics Department, Carnegie Mellon University,
Pittsburgh PA~~ 15213}}
\vspace{.5in}
\baselineskip=24pt
\begin{quotation}
Recent observations of the space distribution of quasars indicate a very
notable
peak in space density at a redshift of $2$ to $3$.It is pointed out in this
article that this may be the result of a phase transition which has a critical
temperature of roughly a few meV (in the cosmological units h=c=k=1 ). It is
further pointed out that such a phase transition is natural in the context of
massive neutrinos. In fact,the neutrino masses required for quasar production
and those required to solve the solar neutrino problem by the MSW mechanism
are consistent with each other.
\end{quotation}
\end{titlepage}
\newpage

\baselineskip=24pt

\vspace{24pt}

It has recently been pointed out by M.Schmidt et al that comoving quasar space
densities exhibit a strong peak at redshifts of $2$ to $3$\re{Scmidt}.In
their article they plot quasar space density as a function of redshift ($z$)
and also separately as a function of cosmic time.There is an unmistakable
peak in the quasar space density in the region of redshift $2$ to $3$.The
authors further point out that the observed decline in space density for
space density for $z>3$ is not a result of instrumentational difficulties
in detecting distant quasars.On the contrary, the decline in observed quasar
density is a real decline in the number density for $z>3$\re{v/vmax}.

Thus,Schmidt et al point out that quasar density peaks sharply at an epoch
of about 2.3 billion years after the Big Bang.The full width at half maximum
of the peak is around 1.4 billion years.Their discussion makes the final
point that one needs to understand these time scales in terms of the formation
and evolution of quasars.

What I would like to argue here is that
such a peaked distribution of objects may result from the formation of
topological defects as the universe goes through a phase transition.Such
topological defects could form the seeds around which quasars light up.

The central power supply of quasars is believed to be gravitational in
origin\re{REES}.It is suggested in this article that this central power supply
may
be formed as a result of topological defect formation in a phase transition
linked with massive neutrinos.Such a phase transition would happen at the right
epoch if one believes the neutrino masses implied by the MSW solution to the
solar neutrino problem.The production of black holes as a result of
cosmological
phase transitions has been discussed by various authors\re{KSS}'\re{BCL}'
\re{Widr}.

The idea of topological defects as seeds for structure formation has been
around for some time.There are many excellent and recent reviews on the
subject\re{Branden}  \re{EdCope}.The idea that topological defects formed
after the decoupling of matter and radiation may play an important role in
structure formation has also been discussed before by Hill,Schramm and
Fry\re{HillFry}(HSF).

Phase transitions that occur after the decoupling of matter and radiation have
been discussed in the literature as Late Time Phase Transitions (LTPT's).The
original motivation for considering LTPTs\re{HillFry} \re{WASS} \re{PRS}
\re{FrHW} \re{GHHK}  \re{HolSing}was the need to reconcile the
extreme isotropy of the Cosmic Microwave Background Radiation (CMBR)\re{CMBR}
with the existence of large scale structure\re{LSS} and also the existence of
quasars at high redshifts\re{quas}.However the hope that LTPT's would not
disturb the CMBR was not realised\re{TurnWW}.However,it may still be that
the CMBR distortions are consistent with the COBE data.
In light of the fact that the
anisotropy measurements of the CMBR are getting better the distortions of the
CMBR predicted by  LTPT's may become important observational tests of these
models.

LTPT's are used to generate density fluctuations either by bubble nucleation
\re{WASS}, dynamics of a slow rolling field\re{PRS} or by the formation of
topological defects such as domain walls(HSF).
The appeal of LTPT's for structure formation lies in the fact that the
density perturbations created are immediately non-linear and so will lead to
the immediate formation of structure.
The models we will be
considering in detail in this article involve the formation of topological
defects as a result of LTPT's.The original idea of the production of soft
domain walls as a result of an LTPT (HSF) has received several criticisms.A
problem with the use of domain walls to form Large Scale Structure has been
pointed out by Kolb and Wang\re{Rockyun}.They argue that domain walls may
never form in a LTPT.However they also point out that domain walls would
form if large fluctuations in the scalar field exist before the phase
transition.
One of the main critcisms of soft domain walls is that they may produce
significant distortions of the CMBR.The most significant microwave distortion
comes from collapsing domain wall bubbles. This has been discussed
and calculated by Turner, Watkins and Widrow (TWW)\re{TurnWW}
.These distortions produce
hot and cold spots on $ \sim 1$ degree angular scales and provide a signature
for the formation of domain walls after recombination.It has been argued that
in the presence of strong friction between the domain walls and the sorrounding
medium the microwave distortions may be significantly reduced\re{Massarotti}.

Since the collapse of closed domain wall bubbles to form black holes which then
act as the central engine of quasars is one of the main themes of this
article we now turn to a quick discussion of the distortions of the CMBR
this produces. As pointed out by TWW this anisotropy is most significant
on $\sim 1^o $ angular scales. The temperature shift due to a photon
traversing a collapsing domain wall bubble is
\be
\frac{\Delta T}{T} = 2.64 \times 10^{-4} h^{-1} \beta A \sigma / (10 MeV^3)
\ee
where h,A,$\beta$ are dimensionless numerical constants of order unity
and $\sigma$ is the surface tension of the domain wall. The present
measurements of the CMBR anisotropy then imply\re{sclu} that
$\sigma < 0.5 MeV^3 $. This constraint, though important to keep in
mind is not a
problem for the viability of the model being presented in this article.
We will return to this point after discussing our particle physics model
for the LTPT.

Discussions of realistic particle physics models capable of generating LTPT's
have been carried out by several authors\re{GHHK}  \re{FrHW}.It has been
pointed
out that the most natural class of models in which to realise the idea of
LTPT's are models of neutrino masses with Pseudo Nambu Goldstone Bosons
(PNGB's).The reason for this is that the mass scales associated with such
models can be related to the neutrino masses,while any tuning that needs to be
done is protected from radiative corrections by the symmetry that gave rise
to the Nambu-Goldstone modes\re{'thooft}.

Holman and Singh\re{HolSing} studied the finite temperature behaviour of the
see-saw model of neutrino masses and found phase transitions in this model
which
result in the formation of topological defects. In fact,the critical
temperature in this model is naturally linked to the neutrino masses.

The motivation for studying the finite temperature behaviour of the see-saw
model of neutrino masses came from a desire to find realistic particle physics
models for Late Time Phase Transitions.

In particle physics one of the standard ways of generating neutrino masses has
been the see-saw mechanism \re{ramslan}.These models involve leptons
and Higgs fields interacting by a Yukawa type interaction.We computed the
finite temperature effective potential of the Higgs fields in this model.An
examination of the manifold of degenerate vacua at different temperatures
allowed us to describe the phase transition and the nature of the topological
defects formed.

To investigate in detail the finite temperature behaviour of the see-saw
model we selected a very specific and extremely simplified version of the
general see-saw model.However,
we expect some of the qualitative features displayed by our specific
simplified model to be at least as rich as those present in more
complicated models.

We chose to study the $2$-family neutrino model.Because of the mass hierarchy
and small neutrino mixings \re{MSWtheory} we hope to capture some of the
essential physics of the $\nuee$-$\numu$ system in this way.The $2$-family
see-saw model we consider requires 2 right handed neutrinos $N_R^i$ which
transform as the fundamental of a global $SU_R(2) $ symmetry.This symmetry is
implemented in the right handed Majorana mass term by the introduction of a
Higgs field $\sigma_{ij}$,transforming as a symmetric rank $2$ tensor under
$SU_R(2)$ (both $N_R^i$ and $\sigma_{ij}$ are singlets under the standard
model gauge group).The spontaneous breaking of $SU_R(2)$ via the vacuum
expectation value (VEV) of $\sigma$ gives rise to the large right handed
Majorana masses required for the see-saw mechanism to work.Also,the spontaneous
breaking of $SU_R(2)$ to $U(1)$ gives rise to 2  Nambu Goldstone Bosons.
The $SU_R(2)$ symmetry is explicitly broken in the Dirac sector of
the neutrino mass matrix,since the standard lepton doublets $l_L$ and the
Higgs doublet $\Phi$ are singlets under $SU_R(2)$.It is this explicit
breaking that gives rise to the potential for the Nambu Goldstone modes
via radiative corrections due to fermion loops.Thus,these modes become Pseudo
Nambu Goldstone Bosons (PNGB's).

The relevant Yukawa couplings in the leptonic sector are:
\be
-{\cal L}_{\rm{yuk}} = y_{ai} \bar{l_L}^a N_R^i \Phi + y \overline{N_R^i}
N_R^{j\ c}
\sigma_{ij} + \rm{h.c.}
\ee
where $a, i, j =1, 2$. The $SU_R(2)$ symmetry is implemented as follows:
\ba
N_R^i & \rightarrow & U^i_j\ N_R^j \nonumber \\
\sigma_{ij} & \rightarrow & U^k_i \sigma_{kl} (U^T)^l_j
\ea
where $U^i_j$ is an $SU_R(2)$ matrix.
The first (Dirac) term breaks the symmetry explicitly.

We can choose the VEV of $\sigma$ to take the form\re{LI}:
$\langle \sigma_{ij} \rangle = f \delta_{ij}$, thus breaking $SU_R(2)$
spontaneously down to the $U(1)$ generated by $\tau_2$ (where $\tau_i$ are the
Pauli matrices).This symmetry breaking gives rise to the $2$ PNGB's,whose
finite temperature effective potential is of interest.

After the Higgs doublet acquires its VEV, we have the following mass terms
for the neutrino fields:
\be
-{\cal L}_{\rm{mass}} = m_{ai} \bar{\nu}_L^a N_R^i + M \overline{N}_R UU^T
N_R^c + \rm{h.c.}
\ee
where $\nu_L^a$ are the standard neutrinos, $m_{ai} = y_{ai}\ v/{\sqrt{2}}$,
$M = y f/\sqrt{2}$.

We can now diagonalize the neutrino mass matrix in the standard see-saw
approximation ($|m_{ai}|<<M$) and perform a chiral rotation to eliminate
the $\gamma_5$ terms.The computation of the complete finite temperature
one-loop effective potential for the PNGB's has been discussed in great
detail in an earlier paper\re{HolSing}.Here we shall only present the results
of that analysis.We will first discuss the simple case where the Dirac mass
matrix $m_{ai}$ is proportional to the identity: $m_{ai} = m \ \delta_{ai}$.
Later we will discuss some trivial modifications to incorporate
the MSW effect into the model.

We parametrized the PNGB's as $\xi_1$ and $\xi_3$ and defined the quantity
$\cal{M}$ to be:
\be
{\cal{M}}^2 = {\frac{m^4}{M^2}} (\cos^2 2 ||\xi|| +
\widehat{\xi}_3^2 \sin^2 2 ||\xi||)
\ee
where $||\xi|| = \sqrt{\xi_1^2 + \xi_3^2}/f$ and
$\widehat{\xi_i} = \xi_i\slash \sqrt{\xi_1^2 + \xi_3^2}$.
Performing the high temperature expansion of the complete potential
and discarding terms of order $(\m^2)^3\slash T^2$ or higher we get,
\be
V_{\rm tot} (\xi_1,\xi_3) = V(\m^2) = (V_0 - \frac{7 \pi^2 T^4}{90}) +
(m_r^2 + T^2/6)\m^2 + \frac{(\m^2)^2}{8 \pi^2}
(n - \log \frac{T^2}{\mu^2}),
\ee
where $n = 2 \gamma - 1 -2 \log \pi \sim -2.1303$,$m_r$ is a parameter in the
model and $\mu$ is the renormalisation scale.$\m$ is naturally of the neutrino
mass scale in this model.

A study of the manifold of degenerate vacua of the effective potential at
different temperatures revealed phase transitions in this model accompanied
by the formation of topological defects at a temperature of a few times the
relevant neutrino mass .Typically at higher temperatures the manifold of
degenerate vacua consisted of a set of disconnected points whereas at lower
temperatures the manifold was a set of connected circles.Thus,domain walls
would
form at higher temperatures which would evolve into cosmic strings at lower
temperatures.

In the Standard Cosmology the redshift range of $2$ to $3$ corresponds to a few
meV in mass scales\re{Rockybook}.Thus the critical temperature of the phase
transition required to produce quasars is fixed at a few meV.

At neutrino detectors around the world, fewer electron neutrinos are
received from the sun than predicted by the Standard Solar Model.
An explanation of the deficiency is offered by the MSW mechanism\re{MSWtheory}
which allows the $\nuee$ produced in solar nuclear reactions to change into
$\numu$.This phenomenon of neutrino mixing requires massive neutrinos with the
masses for the different generations different from each other
\re{MSWtheory}.

The model we considered earlier was an extremely simple one.Although it had 2
families of light neutrinos, there was only one single light neutrino mass.As
such this model was not compatible with the MSW effect.However it is fairly
straightforward to modify our original model to make it compatible with the MSW
effect as is shown in what follows.

To ensure that it is
not possible to choose the  weak interaction eigenstates to coincide with
the mass eigenstates we must require the 2
neutrino mass scales to be different.We can ensure neutrino mixing in our model
by demanding that $m_{ai}$ be such that $m_{11} \ne m_{22}$ and
$m_{12}=0=m_{21}
$.In this case,the effective potential $V_{\rm tot} (\xi_1, \xi_3) =
1/2 (  V( \m _1 ^2) + V( \m _2 ^2) )$ with $ V( \m _i ^2)$ having the same
functional form as$ V( \m^2) (i=1,2)$ and $\m _i^2$ given by the following
expression:
\be
{\m _i}^2 = {\frac{m_{ii}^4}{M^2}} (\cos^2 2 ||\xi|| +
\widehat{\xi}_3^2 \sin^2 2 ||\xi||)
\ee.

Further, if $m_{11}<<m_{22}$  then $V_{\rm tot}(\xi_1, \xi_3) = V(\m _2^2)/2$,
which is exactly half the finite temperature effective potential we discussed
earlier except the neutrino mass scale is the heavier neutrino mass
scale.Hence,
the discussion on phase transitions and formation of topological defects we
carried out earlier goes through exactly except that the critical temperature
is
determined by the mass scale of the heavier of the 2 neutrinos.

In the complete picture of neutrino masses\re{MSWtheory},the neutrinos might
have a mass hierarchy analogous to those of other fermions.Further,we expect
that the mixing between the first and third generation might be particularly
small .
In this
scheme, it is a good first approximation to consider $2$-family mixing.We are
here particularly interested in the $\nuee$-$\numu$ mixing.This is also the
mixing to which the solar neutrino experiments are most sensitive.A complete
exploration of MSW solutions to the solar neutrino problem has recently been
reported by Shi,Schramm and Bahcall\re{MSWexpt}.
We shall restrict ourselves to the $2$-family mixing.
The data seems to imply a central value for the mass of the muon
neutrino to be a few meV\re{Bludman}.

At this point, let us re-examine the constraint on the domain wall tension,
$\sigma$ , placed by the measurements of the CMBR on small angular scales.
Recall that the constraint is $\sigma < 0.5 MeV^3$ . An estimate of
$\sigma$ in terms of the quantities $m_\nu$ and $f$ introduced in our model
can be obtained\re{Widr}.(To make contact with the work of L.Widrow
cited above please note that his $\lambda m^4 = m(\numu)$ and $m = f$
in our notation.) Thus, the constraint on $\sigma$ then implies that
$f < 10^{15} GeV$. Our model is clearly an effective theory with $f$
being some higher symmetry breaking scale on which it is tough to get
an experimental handle. However, the constraint derived above is in fact
natural in the context of the see-saw model of neutrino masses embedded
in Grand Unified Theories as discussed by Mohapatra and Parida (MP)\re{Moha}
and also by Deshpande,Keith and Pal(DKP)\re{Desh}.

There are a few points on which I'd like to add some further comments.The
first has to do with black hole (BH) formation as a result of LTPTs.
The second with the possible role  of a LTPT linked with the mass of
$\nu_{\tau}$ in the formation of LSS (on scales $\sim 100 Mpc$).
Finally,a brief comment on the post COBE status of LTPTs is made.

A number of different groups have studied different mechanisms for black
hole formation as a result of cosmological phase transitions.BH
formation as a result of bubble wall collisions has been suggested by
Hawking et al\re{hawking}.
Collapse of trapped false vacuum domains to produce
BH has been studied by Kodama et al\re{KSS} and Hsu\re{Steve}.The
collapse of closed domain walls  to form BH has been studied by Ipser
and Sikivie\re{sikivie}
and also by Widrow\re{Widr}.In fact, Widrow has suggested that the
collapse of closed domain walls is  likely to produce BH as
a result of LTPTs.An estimate of the mass of BH produced in LTPTs is
$M_{BH} \sim 10^9 M_{solar} (\frac{R}{10^3Mpc})^2 (\frac{f}{10^8GeV}) $,
where $R $ is the radius of the  closed domain wall and we have taken
$m(\numu)$ to be $3 meV$ to obtain our estimate.

There are of course two definite uncertainities in getting a number out
of the above expression.First,there will obviously be a range of
possible sizes of closed domain wall bubbles that will be formed in the
transient period of the phase transition.A natural upper limit to the
size of the bubble will be the horizon size at the epoch of the phase
transitionThus this implies that $R < 10^3 Mpc$. The horizon size
at the onset of the phase transition is  smaller.
Further, it should
be noted that sub-horizon sized bubbles which will be formed in the
transient period of the phase transition may also play an important role.
Secondly,as already pointed out our model is clearly
an effective theory with $f$ being some higher symmetry breaking scale
on which it is tough to get an experimental handle. In fact, values
of $f$ used by DKP and MP give BH masses consistent with observation
within the uncertainities outlined above.

A more detailed analysis of the masses of BH produced in LTPTs as well
as the number densities of these BH will be
the subject of a later work.For now, the fact that it is possible to
produce BH which may act as the central engine of quasars is pointed
out.
One can check that the mechanism of collapse of domain wall bubbles
can in fact give roughly the needed number of black holes to power
quasars. The Hubble radius at the onset of the phase transition
at $ z \sim 5 $ , is a few hundred $ Mpc $. One expects a distribution
of bubble sizes with the horizon as an upper limit to be formed
during a cosmological phase transition. In fact, Turner,
Weinberg and Widrow\re{bubs} have carried out a detailed study
of the distribution of bubble sizes resulting from cosmological
phase transitions. They report that typical bubble sizes in a
successful phase transition range from $0.01$ to $1$ times the
Hubble radius at the epoch of the phase transition and depends
only very weakly on the energy scale of the phase transition.
Thus, the most massive black holes formed from the phase
transition being studied here will have an abundance
$ \sim 10^{-5}$ to $10^{-6} Mpc^{-3} $ with a higher abundance
of less massive black holes. These numbers, in fact match very well
with the observational number densities of quasars as discussed
by Warren and Hewett\re{WH}, and by Boyle et al and Irwin et al\re{BI}.
The point being made is that the epoch of the phase transition
linked with massive neutrinos is about right to explain the observed
peak in the quasar distribution.Further, order of magnitude estimates
of the masses of black holes produced and their abundances
are consistent with observations
within the uncertainities discussed above.A detailed analysis of the
efficiency of black hole formation and therefore an accurate number
for the space density of black holes produced in this model will
be carried out in a later work.

The other point I would like to briefly comment on is the role of a
massive $\nu_\tau$ in the scheme of things presented here.The mass
of $\nu_\tau$ is less well determined but one would expect an
earlier phase transition with a $T_c$ of a few times $m(\nu_\tau)$.
In fact,such a phase transition may well be responsible for the
LSS seen on the scale of $\sim 100 Mpc$.There already exists a lot
of discussion on this subject.

However, BH formation in this earlier phase transition would be more
difficult to observe.In fact,the black holes formed in the most recent
phase transition will have the greatest observable consequences.

The anistropy of the CMBR on large angular scales is more closely
linked to the formation of LSS(on scales $\sim 100$ Mpc).One can
relate the power spectrum of density fluctuations to the gravitational
potential power spectrum responsible for distortions of the CMBR.
This link is thoroughly discussed
by Jaffe,Stebbins and Frieman (JSF)\re{jaffe}.
Since considerable processing of the power spectrum must take place
in the process of black hole formation a relationship between
quasar distribution and the distortions of the CMBR on large angular
scales is considerably more difficult to establish.
The viability of LTPTs in the post-COBE era has recently been discussed by
Schramm and Luo\re{sclu}.They point out that LTPTs are still a viable
model for the formation of LSS.

This paper is devoted to the subject of quasar formation. To place things
in perspective one should keep in mind that different mechanisms may play
important roles in structure formation at different length scales.This
point has been emphasised by Carr\re{carr}who has given a comprehensive
discusssion on the origin of cosmological density fluctuations.
In fact, JSF  also conclude their discussion by pointing out that the final
power spectrum is most likely due to a combination of primordial and late
time effects.

In conclusion,the MSW solution to the solar neutrino problem seems to imply a
muon
neutrino mass of a few meV.This in turn would lead to a phase transition in
the PNGB fields associated with massive neutrinos with a critical temperature
of several meV.This phase transition happens at the correct epoch in the
evolution of the universe to provide a possible explanation of the peak in
quasar space density at redshifts of $2$ to $3$.This article is clearly
only a first step in bringing these ideas together.There are clearly
some issues which need to be addressed in greater depth and explored in fuller
detail. Thus, a more detailed and accurate analysis of the black hole
formation efficiency and the masses and
number densities of BH's needs to be carried
out.

To obtain more precise estimates of the black hole number densities and
masses one needs to make a more detailed analysis of domain formation
and growth in FRW cosmologies.Domain formation and growth are among the
most important non-equilibrium phenomena associated with phase transitions.
The real time formulation of finite-temperature field theory is the most
logical and well-suited formalism to study such phenomena.

Boyanovsky, Lee and Singh have studied
domain formation and growth and discussed it at length using
the real time time formalism\re{boysinlee}.This study was carried out in
Minkowski space. We need to extend this work to FRW cosmologies.
In fact, non-equilibrium phenomena such as particle creation, entropy
growth and dissipation have already been studied for FRW cosmologies
by Boyanovsky,de Vega and Holman\re{boyvh}.We now need to examine
domain formation and growth in this formalism.

Work in this direction has already been started. In fact, Holman and
Singh are presently investigating the growth and formation of axion
domain walls near the QCD phase transition\re{holsinax}.This work
among other things, is setting up the formalism to study domain formation
and growth in FRW cosmologies.This formalism will then be used to study
domain formation and growth in Late Time Phase Transitions in general.
In particular, it will make more precise estimates of the number densities
and masses of black holes produced in LTPTs.

This number density and mass distribution of BH's then needs
to be compared to the observational data on quasars.Further, a more
detailed
analysis of the angular dependence of the CMBR anisotropy produced
needs to be carried out and compared to observations. I shall be addressing
these issues in a later work on the subject.
 My hope is
that this work will stimulate further thoughts about these issues.

\centerline{\bf ACKNOWLEDGEMENTS}

I would like to thank the organisors of Texas/PASCOS '92 for giving me the
opportunity to attend an extremely stimulating conference.I would like to thank
Maarten Schmidt for a lucid talk on the subject of Quasar Distributions and
Roger Blandford for helping to remove the remaining doubts.It's a pleasure
to thank Richard Holman for many valuable discussions,critical comments and
helpful suggestions.For helpful discussions on different aspects of this
work I thank Lincoln Wolfenstein, Cyril Hazard, David Turnshek and
Valery Khersonsky.I would like to thank the referees for helping me to
focus on the issues that need further in depth exploration.
This work was supported in part by the DOE contract DE--FG02--91ER--40682.

\frenchspacing

\newpage

\vspace{36pt}

\centerline{\bf REFERENCES}

\frenchspacing

\begin{enumerate}

\item\label{Scmidt} M.Schmidt,D.P.Schneider,J.E.Gunn in The Space Distribution
of Quasars, ASP Conference Series,Vol.21,1991,David Crampton(ed.),p.109 .

\item\label{v/vmax} This point is discussed in detail in the section on
$V/V_{max}$ of the Scmidt et al article quoted above.

\item\label{KSS} H.Kodama,M.Sasaki,K.Sato, Prog. Theor. Phys. 68,
1979 (1982);

\item\label{BCL}J.D.Barrow,E.J.Copeland,A.R.Liddle Phys. Rev. D46, 645 (1992).

\item\label{Widr} L.M.Widrow Phys. Rev.D40,1002 (1989).

\item\label{REES} M.J.Rees in Quasars:Proc. 119th Symposium of the IAU,G.Swarup
and V.K.Kapahi (eds), 1 (1985).

\item\label{Branden} R.H.Brandenberger,Brown-HET-881,September 1992.

\item\label{EdCope} E.Copeland, The Physical Universe:The interface

between Cosmology,Astrophysics and Particle Physics, Proceedings,

Lisbon, Portugal, 1990, J.D.Barrow,A.B.Henriques,M.T.V.T. Lago

and M.S.Longair (Eds.)

\item\label{HillFry} C.T.Hill,D.N.Schramm, and J.Fry,Comm. on Nucl. and Part.
Phys. 19, 25 (1989).

\item\label{CMBR} G.F.Smoot, et al., \ApJ {\bf 360}, 685 (1991).

\item\label{LSS} M.Geller and J.Huchra, {\em Science} {\bf 246}, 897
(1989);

A.Dressler, S.M.Faber, D.Burstein, R.L.Davies, D.Lynden-Bell,
R.J.Turlevich, and G.Wegner, \ApJ {\bf 313}, L37 (1987);

V. de Lapparent, M.J.Geller, and J.P.Huchra, \ApJ {\bf 302}, L1 (1986);

J.R.Gott III, A.L.Melott, and M.Dickinson,\ApJ {\bf 306}, 341 (1986);

M.P.Haynes and R.Giovanelli, \ApJ {\bf 306}, L55 (1986).

\item\label{quas} D.P.Schneider,M.Scmidt,J.E.Gunn, Institute for Advanced
Study Report IASSNS--AST 91/30 (unpublished).

\item\label{TurnWW} M.S.Turner,R.Watkins,L.M.Widrow
\ApJ {\bf 367}, L43-47, (1991)

\item\label{WASS} I.Wasserman, {\em Phys. Rev. Lett.} {\bf 57}, 2234 (1986).

\item\label{PRS} W.Press, B.Ryden, D.N.Spergel, {\em Phys. Rev. Lett.}
{\bf 64}, 1084 (1990).

\item\label{Rockyun} E.~W.~Kolb, and Y.~Wang, "Domain Wall Formation in late-
time phase transitions", University of Florida Institute for Fundamental Theory
preprint UFIFT-HEP-92-5, (January 1992).(submitted to Phys. Rev. D)

\item\label{Massarotti} A.Massarotti, Phys. Rev. {\bf D43}, 346, (1991).

\item\label{FrHW} J.A.Frieman,C.T.Hill,R.Watkins, Phys. Rev {\bf D46}, 1226,
(1992).

\item\label{GHHK} A.K.Gupta,C.T.Hill,R.Holman and E.W.Kolb, Phys. Rev. D45, 441
(1992).

\item\label{'thooft} see, e.g.,G. 'tHooft  "Naturalness, Chiral Symmetry,and
Spontaneous Chiral
Symmetry Breaking," in {\em Recent Developments in Gauge
Theories}, (Plenum Press, New York, 1980), G. 't Hooft, et al., eds.

\item\label{ramslan} M.Gell-Mann,P.Ramond,R.Slansky, in Supergravity,North
Holland Pub.(1979);
T.Yanagida, in "Proc. of the Workshop on Unified Theories and Baryon Number in
the Universe," KEK,Japan (1979).

\item\label{HolSing} R.Holman,A.Singh, Phys. Rev. D, Jan.15,1992.

\item\label{LI} L.F.Li, Phys.Rev. D9, 1723 (1974)

\item\label{Rockybook} E.W.Kolb and M.S.Turner:The Early Universe,73.(Addison-
Wesley Pub. Co., 1990).

\item\label{MSWtheory} L.Wolfenstein,CMU-HEP88-14 and Proc. 1988 SLAC Summer
Institute-July 18-29, 1988.

\item\label{MSWexpt}   X.Shi,D.N.Schramm,J.N.Bahcall,{\em Phys. Rev. Letters},
{\bf 69}, 717, (1992).

\item\label{Bludman}  S.A.Bludman, N.Hata, D.C.Kennedy, and Paul Langacker,
UPR-0516T(revised), "Implications of Combined Solar Neutrino Observations
and Their Theoretical Uncertainities".

\item\label{Moha} Mohapatra,R.N. and Parida,M.K. , Phys. Rev. D, {\bf47},
264 (1993)

\item\label{Desh} Deshpande,N.G. ,Keith, E., and Pal,P.B. , Phys. Rev D,
{\bf46} ,2261 (1992).

\item\label{hawking} S.W.Hawking, I.G.Moss and J.M.Stewart, Phys. Rev. D,
30, 2681,(1982).

\item\label{Steve} S.D.H.Hsu, Phys. Lett. B, 251, 343, (1990).

\item\label{sikivie} J.Ipser and P.Sikivie, Phys. Rev. D, 30, 712, (1984).

\item\label{bubs} M.S.Turner , E.J.Weinberg and L.M.Widrow, Phys.
Rev. D, 46, 2384 (1992).

\item\label{WH} S.J.Warren and P.C.Hewett, Reports on Progress in Physics,
53, (Aug. 1990), pp.1095-1135. In particular, please refer to figs. 3,4
and 7 on p.1105, 1109 and 1123 respectively.

\item\label{BI} B.J.Boyle, L.R.Jones, T.Shanks, B.Marano, V.Zitelli
and G.Zamorani in The Space Distributrion of Quasars,(Ed. D.Crampton),
ASP Conference Series, Vol. 21 (June 1991), pp.193;

M.Irwin, R.G.McMahon, C.Hazard {\it ibid.} pp. 122 .

\item\label{jaffe}A.H.Jaffe,A.Stebbins and J.A.Frieman, Minimal Microwave
Anisotropy from Perturbations Induced at Late Times.
Fermilab-Pub-92/362-A.

\item\label{sclu}D.N.Schramm and X.Luo, The Phenomenological Status of
Late Time Phase Transitions after COBE,
Fermilab-Pub-93/020-A.

\item\label{carr} B.J.Carr , Nuclear Physics B252 (1985), 81-112.

\item\label{boysinlee} D.Boyanovsky,D-S.Lee and A.Singh, Phys. Rev. D48
(July 1992), 800-815.

\item\label{boyvh} D.Boyanovsky,H.J. de Vega and R.Holman, Non-equilibrium
evolution of scalar fields in FRW Cosmologies I , PITT-93-6,LPTHE-93-52,
CMU-HEP-93-21,DOE-ER/40682-46.

\item\label{holsinax} R.Holman and A.Singh, work in progress.

\end{enumerate}

\newpage

\vspace{36pt}

\end{document}